\documentclass[preprint,showpacs,preprintnumbers,amsmath,amssymb,superscriptaddress]{revtex4}
\usepackage{graphicx}
\usepackage{amsmath}
\usepackage{amssymb}
\usepackage{epsfig}
\usepackage{amsthm}
\usepackage{color}
\def\be{\begin{equation}}	
\def\ee{\end{equation}}
\def\arr{\begin{array}{rll}}
\def\ea{\end{array}}
\def\bea{\begin{eqnarray}}
\def\eea{\end{eqnarray}}
\begin{document}

\title{Discrete spectrum Radiation from a charged particle moving in a medium with Maxwell fish eye refraction index profile}

\author{Zhyrair Gevorkian}
\affiliation{Yerevan Physics Institute, Alikhanian Brothers St. 2, 0036, Yerevan, Armenia}
\affiliation{Institute of Radiophysics and Electronics, Ashtarak-2,0203, Armenia }
\author{Mher Davtyan}
\affiliation{Institute of Radiophysics and Electronics, Ashtarak-2,0203, Armenia }
%\author{Armen Allahverdyan}
%\affiliation{Yerevan Physics Institute, Alikhanian Brothers St. 2, 0036, Yerevan, Armenia}
%\affiliation{Bogoliubov  Laboratory of Theoretical Physics, JINR,
%141980 Dubna, Russia}
%\affiliation{Yerevan State University, 1 Alex Manoogian St., Yerevan, 0025, Armenia}
\begin{abstract}
Radiation from a charged particle moving in a medium with Maxwell fish eye refraction index profile is considered. It is shown that the radiation spectrum has a discrete character. The main emitted wavelength is proportional to the refractive profile's radius and has a dipole character in a regular medium. Cherenkov like threshold velocity is established. A cardinal rearrangement of angular distribution in a lossless medium is predicted. This behavior is caused by the total internal reflection in a lossless medium as apposed to photons' attenuated total  reflection in the regular medium. Lossless medium ensures that both directed and monochromatic emission can serve as a light source in the corresponding regions.
\end{abstract}
\pacs{41.60-m,77.22.Ch,84.40.-x}
\maketitle

\section{Introduction}
Recently the interest in the Maxwell fish eye \cite{bowolf} refraction index profile has increased dramatically. The reasons are its possible use in cloaking phenomena \cite{Leonhardt06}, perfect imaging \cite{Leonhardt09,Philbin10,Blaikie10,pazynin15}, quantum optics with single atoms and photons \cite{perczel2018}, optical resonators \cite{turks14,gevdav20}, etc. Earlier we have shown \cite{gevdav20} that apart from the spherical symmetry, Maxwell fish eye possesses an additional symmetry also. The extended symmetry leads to additional integrals of motion. In the geometrical optics limit, all the photon trajectories are closed and their parameters are expressed through the integrals of motion \cite{gevdav20}.
To detect a cloak in the Maxwell fish eye medium, it was suggested \cite{singap} to use the radiation induced by the motion of a charged particle.
Consideration was realized by exploring the dyadic Green's functions \cite{dyadic}. It was revealed that the emitted radiation is a mix of Cherenkov and transition radiations.

 As apposed to the research regarding light propagation in Maxwell fish eye medium, much less attention has been paid to light generation problems in that particular medium. However, it turns out that radiation emitted by a charged particle when it passes through such a medium possesses unique properties as well (see below).
In the present paper, we consider the spectrum and angular distribution of radiation from a charged particle moving in a Maxwell fish eye refraction profile medium. Instead of dyadic Green's function, we utilize the exact Green's function of scalar Helmholtz equation \cite{demkov71,poland,pazynin}. This approach allows us to obtain complete analytical expressions for radiation intensity that reveal new physical results.

\section{Initial Relations}
We  start from the Maxwell equations for the field Fourier components
\begin{eqnarray}
 {\bf \nabla\times E(r,\omega)}=\frac{i\omega}{c}{\bf B(r,\omega)}, \quad {\bf \nabla\times H(r,\omega)}=\frac{4\pi}{c}{\bf j(r,\omega)}-\frac{i\omega}{c}{\bf D(r,\omega)}, \nonumber \\
 {\bf \nabla\cdot D}=4\pi \rho({\bf r},\omega),\quad {\bf \nabla\cdot B}=0,\quad {\bf B(r,\omega)}=\mu({\bf r},\omega) {\bf H(r,\omega)},\quad {\bf D(r,\omega)}=\varepsilon({\bf r},\omega){\bf E(r,\omega)}
\label{maxwell}
\end{eqnarray}
where $\rho$ and ${\bf j}$ are the charge and current densities associated with the moving particle.  We proceed with the calculations using vector potential ${\bf A}$ and scalar potential $\phi$ instead of ${\bf E}$ and ${\bf B}$. From Eq.(\ref{maxwell}) those can be introduced in the following way
\begin{equation}
{\bf B}={\bf \nabla\times A},\quad {\bf E}-\frac{i\omega}{c}{\bf A}={\bf \nabla}\phi
\label{potential}
\end{equation}
 By substituting the expressions for $E$ and $B$ into the second equation in Eq.(\ref{maxwell}), we get to the following equation
\begin{equation}
{\bf\nabla}\times\frac{1}{\mu}{\bf\nabla}\times{\bf A}=\frac{4\pi}{c}{\bf j}-\frac{i\omega}{c}\varepsilon({\bf\nabla}\phi+\frac{i\omega}{c}{\bf A})
\label{mixed}
\end{equation}
From Eq.(\ref{mixed}), assuming that $\mu({\bf r},\omega)$ is a slowly varying function in the space and using the property of double curl, we obtain
\begin{equation}
\nabla^2{\bf A}+\frac{\omega^2}{c^2}\varepsilon({\bf r})\mu({\bf r}){\bf A}=-\frac{4\pi}{c}\mu{\bf j}+{\bf\nabla}(\nabla\cdot{\bf A})+\frac{i\omega}{c}\varepsilon({\bf r})\mu({\bf r}) {\bf \nabla}\phi
\label{mixed2}
\end{equation}
%Earlier, for consideration of radiation in Maxwell fish eye profile \cite{singap}, the dyadic Green's functions \cite{dyadic} were used. Here %we use another approach namely  the Green's function of scalar Helmholtz equation which for Maxwell fish eye profile is known %\cite{demkov71,poland,pazynin}  as well.
 We take the gauge condition for inhomogeneous medium in the the following form
\begin{equation}
{\bf\nabla}(\nabla\cdot{\bf A})+\frac{i\omega}{c}\varepsilon({\bf r})\mu({\bf r}) {\bf \nabla}\phi=0
\label{gauge}
\end{equation}
and eventually find the equation for the vector potential
\begin{equation}
\nabla^2{\bf A}+\frac{\omega^2}{c^2}\varepsilon({\bf r})\mu({\bf r}){\bf A}=-\frac{4\pi}{c}\mu{\bf j}
\label{waveA}
\end{equation}
 Note that although the condition (\ref{gauge}) is similar to the Lorenz gauge condition for inhomogeneous media, it does not result to completely decoupled equations for vector and scalar potentials. However, our choice of the gauge condition leads to a less complex equation for the vector potential, which is more important when examining problems regardings radiation.
 Note also that when deriving Eq.(\ref{waveA}) we assume the slow variance only for $\mu({\bf r})$ but not for $\varepsilon({\bf r})$. This means that our consideration is correct for quite large class of materials (particularly all nonmagnetic mediums $\mu=1$). Also note that the form of equation for ${\bf A}$ can be changed depending on the gauge condition we choose \cite{bound}. The point is that in an inhomogeneous medium Lorenz gauge condition acquires different forms \cite{bound}. Again here we choose the one that leads to the simplest equation for the vector potential.

It follows from the Eq.(\ref{waveA}) that  the radiation vector potential associated with the external source is directed similar to current density ${\bf j}$ which is assumed to be directed along $z$, see Fig.1. Therefore the radiation potential can be expressed through the Green's function of scalar Helmholtz equation
\begin{equation}
A_{zr}({\bf R})=-\frac{4\pi}{c}\int d{\bf r}G({\bf R,r})\mu({\bf r})j_z({\bf r}).
\label{radfield}
\end{equation}
The latter equation represents a particular solution of Eq.(\ref{waveA}) associated with an external source. To obtain the general solution, one should add also the solution of homogeneous equation. However, when examining the far field radiation, it is the particular solution Eq.(\ref{radfield}) that gives the main contribution.
Green's function satisfies the equation
\begin{equation}
[\nabla^2+\frac{\omega^2}{c^2}n^2({\bf R})]G({\bf R,r})=\delta({\bf R}-{\bf r})
\label{grfunc}
\end{equation}
where  $n({\bf r})=\sqrt{\varepsilon({\bf r})\mu({\bf r}})$ is the refraction index of the medium. We choose it in the form (see Fig.1)
\begin{eqnarray}
n(r)=\frac{2n_0\rho^2}{r^2+\rho^2},\quad r<R_1 \nonumber\\
      1,\quad r>R_1
\label{fisheye}
\end{eqnarray}

\begin{figure}
\begin{center}
\includegraphics[width=8.0cm]{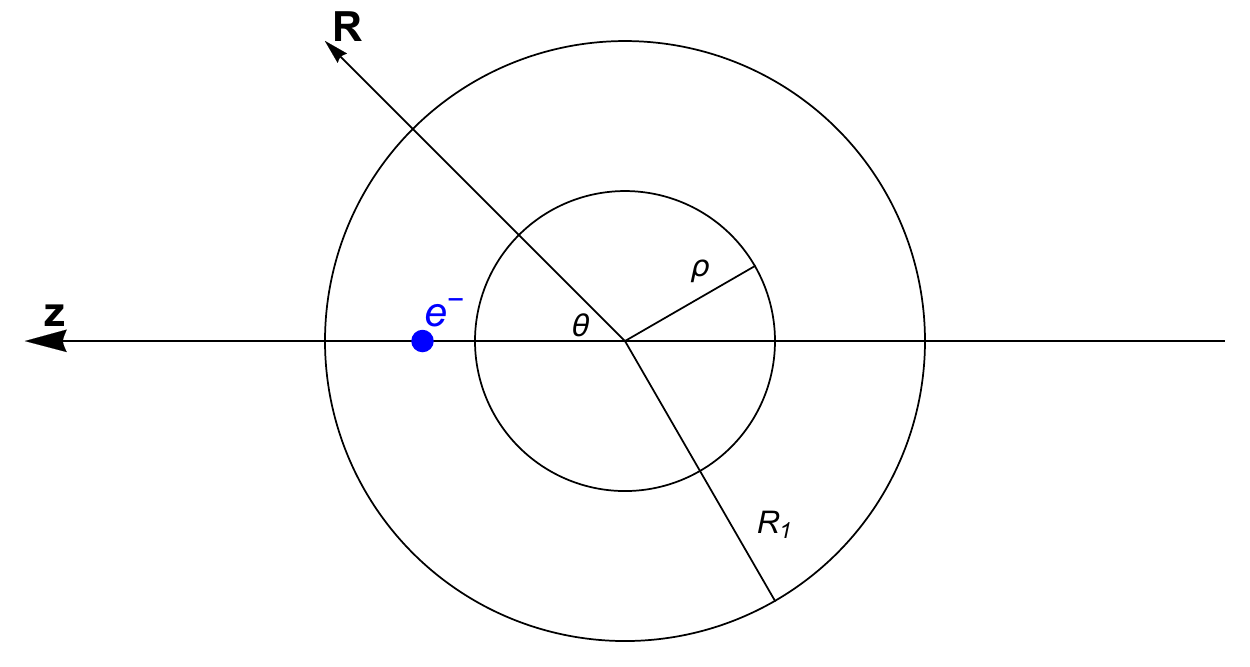}
\caption{Geometry of the problem. Observation point is far away from the charge and from the core of the refraction profile.}
\label{fig.1}
\end{center}
\end{figure}

Green's problem Eq.(\ref{grfunc}) is exactly solved \cite{demkov71,poland}
\begin{eqnarray}
G_{\nu}({\bf R,r})=-\frac{1}{4\pi\cos(\pi\nu)}\frac{\sqrt{(R^2+\rho^2)(r^2+\rho^2)}}{|{\bf R-r}|\sqrt{R^2r^2+2\rho^2{\bf Rr}+\rho^4}}\times \nonumber\\\times \sin\left[(\nu+1/2)\arccos\left(-1+\frac{2\rho^2({\bf R-r})^2}{(R^2+\rho^2)(r^2+\rho^2)}\right)\right]
\label{grfunsol}
\end{eqnarray}
where
\begin{equation}
\nu=\frac{-1+\sqrt{1+4n_0^2k^2\rho^2}}{2}.
\label{nu}
\end{equation}
Here $k=\omega/c$ and $\nu\neq m+1/2, -m-3/2$ $(m\in \textrm{N})$. At these specific values, as it is seen from Eq.(\ref{grfunsol}), Green's function is divergent and one needs another expression \cite{poland}
\begin{eqnarray}
\tilde{G}_{m+1/2}({\bf R,r})=(-)^m\frac{1}{4\pi^2}\frac{\sqrt{(R^2+\rho^2)(r^2+\rho^2)}}{|{\bf R-r}|\sqrt{R^2r^2+2\rho^2{\bf Rr}+\rho^4}}\times\nonumber\\  \times\left\{\cos\left[(m+1)\arccos\left(-1+\frac{2\rho^2({\bf R-r})^2}{(R^2+\rho^2)(r^2+\rho^2)}\right)\right]\times \right.\nonumber\\ \left.\times\arccos\left(-1+\frac{2\rho^2({\bf R-r})^2}{(R^2+\rho^2)(r^2+\rho^2)}\right)+\right.\nonumber\\
\left. +\frac{\sin\left[(m+1)\arccos\left(-1+\frac{2\rho^2({\bf R-r})^2}{(R^2+\rho^2)(r^2+\rho^2)}\right)\right]}{2(m+1)}\right\}
\label{grfuncother}
\end{eqnarray}

The current density corresponding to the particle with charge $e$  moving along the $z$ axis with velocity $v$ has the form  ${\bf j}({\bf r}, t)=e{\bf v}\delta(x)\delta(y)\delta(z-v_z t)$. The corresponding Fourier component which determines the Fourier component of the radiation vector potential will have the following form
\begin{equation}
{\bf j}({\bf r},\omega)=\frac{e{\bf v}}{v}\delta(x)\delta(y)e^{ik_0z}
\label{chargecurr}
\end{equation}
where $k_0=\omega/v$. Using the expressions for Green's functions Eqs.(\ref{grfunsol},\ref{grfuncother}) and the expression for the current density Eq.(\ref{chargecurr}), one can find radiation potential and radiation intensity.
\section{radiation potential}
First, we determine the radiation potential in the region $R<R_1$
\begin{equation}
A({\bf R})=-\frac{4\pi}{c}\int_{r<R_1}d{\bf r}G({\bf R,r})\mu({\bf r})j({\bf r}),\quad R<R_1
\label{vecpot}
\end{equation}
Here $A\equiv A_z, j\equiv j_z$ .
According to the Green's theorem, in Eq.(\ref{vecpot}) it should also appear an additional surface integral over the sphere of radius $R_1$. However, for large $R_1\gg \rho$, this surface term falls  faster than $1/R_1$, therefore it does not give a contribution to the radiation potential. As it seen from Eq.(\ref{grfunsol}) Green's function has singularities at the points $\nu=m+1/2$. In order to overcome this difficulty we assume a small imaginary part for $n_0$ and correspondingly for $\nu$. As it is noted in \cite{poland} the expression Eq.(\ref{grfunsol}) is correct for complex values of $\nu$  either.  For these discrete values $\nu$ the integral Eq.(\ref{vecpot}) can be calculated analytically \cite{ryzhik}. We present the results for $\nu=1/2+iIm[\nu_{1/2}]$ and $\nu=3/2+iIm[\nu_{3/2}]$, $Im[\nu]\ll 1$
\begin{eqnarray}
A_{1/2}(R)=-\frac{4e}{c}\frac{i sign(v)K_0(k_0\rho)}{\sinh(\pi Im[\nu_{1/2}])}\frac{\rho}{\sqrt{R^2+\rho^2}},\nonumber\\
A_{3/2}(R)=-\frac{8ie sign(v)}{c \sinh(\pi Im[\nu_{3/2}])} \frac{\rho}{\sqrt{R^2+\rho^2}}\left[(k_0\rho)K_0(k_0\rho)-K_1(k_0\rho)\right],\quad R<R_1
\label{vecpotspec}
\end{eqnarray}
where $K_{0,1}$ are the modified Bessel functions of second kind \cite{ryzhik} and non-magnetic medium ($\mu\equiv 1$) is assumed (see below).
When obtaining Eq.(\ref{vecpotspec}) in the limit $R,R_1\gg \rho$ we extend the integral limits in Eq.(\ref{vecpot}) to infinity and neglect all the terms smaller $1/R$. Note that radiation potential depends only on the module of the vector ${\bf R}$. Non isotropic terms are possible for non-discrete values of $\nu$, however they are small in terms of the parameter $1/R$. In order to find the radiation intensity, one should know the radiation fields in the vacuum region that matches the solutions given in Eq.(\ref{vecpotspec}).

%In the vacuum region in order to match the isotropic vector potentials in the medium Eq.(\ref{vecpotspec}),
In the absence of external sources, the isotropic solution of Eq.(\ref{waveA}) is chosen in the form
\begin{equation}
A_v(R)=C\frac{e^{ikR}}{R},\quad R>R_1
\label{vac}
\end{equation}
The constant $C$ should be found from the boundary conditions. Since there is no any current on the sphere $R_1$ (${\bf R_1}\neq \hat{\bf z}R_1$), the magnetic field is finite and correspondingly we have (see for example \cite{bound})
\begin{equation}
{\bf nxA_1=nxA_2}
\label{bound}
\end{equation}
where ${\bf n}$ is unit vector normal to the boundary surface. In our case, this leads to the equation $A_v(R_1)=A_{1/2}(R_1)$. Using Eqs.(\ref{vecpotspec},\ref{vac}) one finds
\begin{equation}
C_{1/2}(R_1)=-\frac{4iesign(v)}{c\sinh(\pi Im[\nu_{1/2}])}\rho K_0(k_0\rho)e^{-ikR_1}
\label{constant}
\end{equation}
When obtaining Eq.(\ref{constant}) we assumed that $R_1\gg\rho$. One can also find $C_{3/2}$ and other values of $C$ for $Re[\nu]=m+1/2$ in analogous manner. For the non-discrete values of $\nu$, $C$ can be found numerically (see Fig.2).

\begin{figure}
\begin{center}
\includegraphics[width=8.0cm]{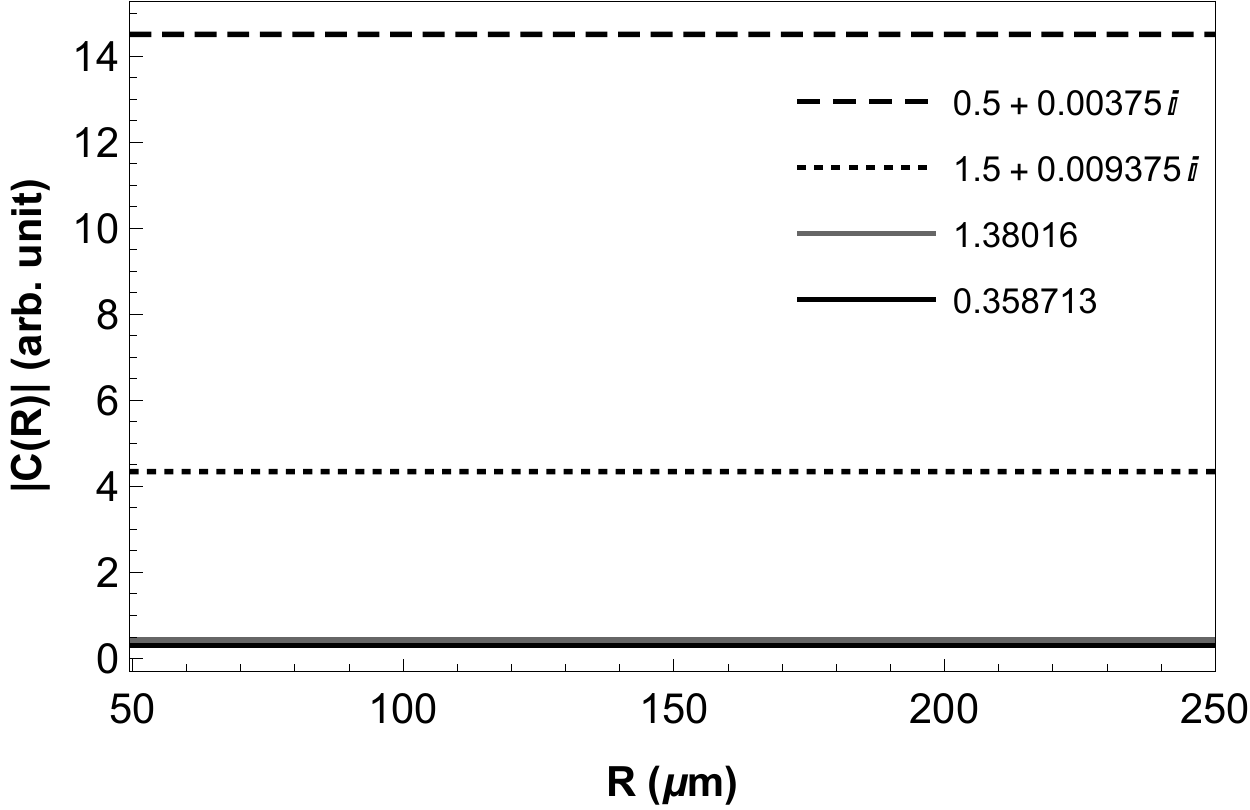}
\caption{$|C(R)|$ dependence on $R$ for different values of $\nu$ when $n_0 = 5, \beta = 0.9$. Straight lines mean that $|A(R)|\sim 1/R$. The amplitude of the radiation is negligible for $\nu \neq m+1/2$.}
\label{fig.2}
\end{center}
\end{figure}
As it follows from Fig.2 one can expect any significant radiation emission only at frequencies for which $\nu=m+1/2$.

\section{radiation intensity}
To find the radiation intensity, one should know the electric and magnetic fields far away from the charge outside of the medium area $R>R_1\gg\rho$ (see Fig.1).
Recall that the radiation vector potential is directed along the $z$ axis and ${\bf A}\equiv A_z$. Therefore from Eq.(\ref{vac}) the magnetic field ${\bf B}={\bf \nabla\times A}$ at the observation point $R$ is given by expressions

\begin{equation}
B_x({\bf R})=C\frac{ikR_ye^{ikR}}{R^2},\quad B_y({\bf R})=-C\frac{ikR_xe^{ikR}}{R^2},\quad B_z\equiv 0
\label{magfield}
\end{equation}

Here we keep only the terms which are proportional to $O(1/R)$ that give contribution to the radiation intensity. The magnetic field energy density follows from Eq.(\ref{magfield})
\begin{equation}
U_B=\frac{|{\bf B}|^2}{8\pi}=|C|^2\frac{k^2\sin^2\theta}{8\pi R^2}
\label{mageng}
\end{equation}
The electric field energy density in the vacuum is equal to the magnetic field energy density and the radiation intensity is determined through the electromagnetic field energy density as $I(\theta)=cR^2U$, where $U=2U_B$. Using Eqs.(\ref{constant},\ref{magfield}) for the radiation intensity
at $Re[\nu]=1/2$, we have
\begin{equation}
I_{1/2}(\theta)=\frac{4e^2}{\pi c}\frac{k^2\rho^2K_0^2(k_0\rho)}{\sinh^2(\pi Im[\nu_{1/2}])}sin^2\theta
\label{inten}
\end{equation}
It follows from Eq.(\ref{nu}) that for $n_0^2=\varepsilon+iIm[\varepsilon],\quad Im[\varepsilon]\ll \varepsilon$ and $\nu=1/2+iIm[\nu_{1/2}]$
\begin{equation}
k\rho=\frac{\sqrt{3}}{2\sqrt{\varepsilon}},\quad Im[\nu_{1/2}]=\frac{3Im[\varepsilon]}{8\varepsilon}
\label{spcase}
\end{equation}

Substituting Eq.(\ref{spcase}) into Eq.(\ref{inten}),we finally obtain
\begin{equation}
I_{1/2}(\theta)=\frac{3e^2}{\pi c}\frac{K_0^2(\frac{\sqrt{3}}{2\beta\sqrt{\varepsilon}})}{\sinh^2(\frac{3\pi Im[\varepsilon]}{8\varepsilon})}sin^2\theta
\label{final}
\end{equation}
where $k=k_0\beta$ and $\beta=v/c$. Here we present a peak intensity corresponding to the wavelength $\lambda=4\pi\sqrt{\varepsilon/3}\rho$ $(Re[\nu]=1/2)$.  Similar expressions can be written for smaller peak intensities $Re[\nu]=3/2$, etc. Intensities for non-discrete frequencies are significantly smaller (see Fig.1). It is well known \cite{ryzhik} that modified Bessel function $K_0$ is exponentially small for large values of the argument. Therefore from Eq.(\ref{final}) we can state that for the existence of radiation, the following condition should be satisfied
\begin{equation}
\beta>\frac{\sqrt{3}}{2\sqrt{\varepsilon}}
\label{cheren}
\end{equation}
This is the analogue of Cherenkov condition \cite{mik72} for the Maxwell fish eye profile. Note that the radiation considered here is the mix of Cherenkov and transition radiations, see also \cite{singap}. As it follows from Eq.(\ref{cheren}), radiation emission condition in Maxwell fish eye profile is weaker than the ordinary Cherenkov condition in the homogeneous medium  with refraction index $\sqrt{\varepsilon}$, $\beta>1/\sqrt{\varepsilon}$. However, it is stronger than the Cherenkov condition for the homogeneous medium with refraction index $2\sqrt{\varepsilon}$. It is interesting that condition obtained for totally inhomogeneous medium is very similar to Cherenkov condition for the homogeneous medium.

 As it seen from Eq.(\ref{final}), the angular distribution of the intensity is like that of dipole radiation. Moreover, the maximum intensity is reached in the directions normal to the particle velocity.
%Note that the radiation intensity does not depend on the cutting distance $R_1$. This is the result of the fact that radiation fields behave as $1/R$ %and correspondingly $|C|$ does not depend on $R_1$, see Eq.(\ref{constant}). However this is not always the case and in the lossless medium %$Im[\varepsilon]=0$ the situation is changed(see below).

At the end of this paragraph we present radiation intensity for the impedance match medium $\mu(r)=\varepsilon(r)=n(r)$ (see \cite{Philbin10})
\begin{equation}
I_{1/2}^i(\theta)=\frac{9e^2}{\pi c}\frac{K_1^2(\frac{\sqrt{3}}{2\beta\sqrt{\varepsilon}})}{\varepsilon\beta^2\sinh^2(\frac{3\pi Im\varepsilon}{8\varepsilon})}sin^2\theta
\label{imp}
\end{equation}

Comparing with the nonmagnetic medium case Eq.(\ref{final}) one can see that the main difference is the additional particle energy dependence ($\sim 1/\beta^2$).
\subsection{Lossless medium. Non isotropic radiation}
The main difference is happening at the discrete frequencies $\nu=m+1/2$. In this case, the radiation potential is determined through the generalized Green's function Eq.(\ref{grfuncother}). Our estimates show that the radiation potential at large distances $R\gg \rho$ behaves
as
\begin{eqnarray}
\tilde{A}_{1/2}({\bf R})&\sim \frac{8esign(v)}{\pi c}\int_{-R_1}^{R_1} dz\frac{(z^2-\rho^2-4\rho^2 z\cos\theta/R)e^{ik_0z}}{\sqrt{[(z-R\cos\theta)^2+R^2\sin^2\theta](z^2+\rho^2)[(z+\rho^2\cos\theta/R)^2+\rho^4\sin^2\theta/R^2]}} \nonumber\\
&\arccos\frac{\rho^2-z^2-4\rho^2 z\cos\theta/R}{z^2+\rho^2}
\label{anis}
\end{eqnarray}
 Unfortunately this integral can not be taken analytically and we are forced to use some numerical estimations. Nevertheless we will try to collate analytic and numerical results and make qualitative predictions on radiation properties in different cases.
The second term in Eq.(\ref{grfuncother}) for $Re[\nu]=1/2$ has an isotropic character and is analogous to the regular case in Eq.(\ref{grfunsol}). It is obvious that at small angles $\theta\to 0$, the main contribution to the integral Eq.(\ref{anis}) gives the pole at $z=-\rho^2\cos\theta/R$. Conversely, the contribution of the pole $z=R\cos\theta$ at large distances is negligible because of the oscillations $e^{ik_0z}$ in the integral.
We present the results of numerical estimates of the integral Eq.(\ref{anis}) in Fig.3,4.
\begin{figure}
\begin{center}
\includegraphics[width=8.0cm]{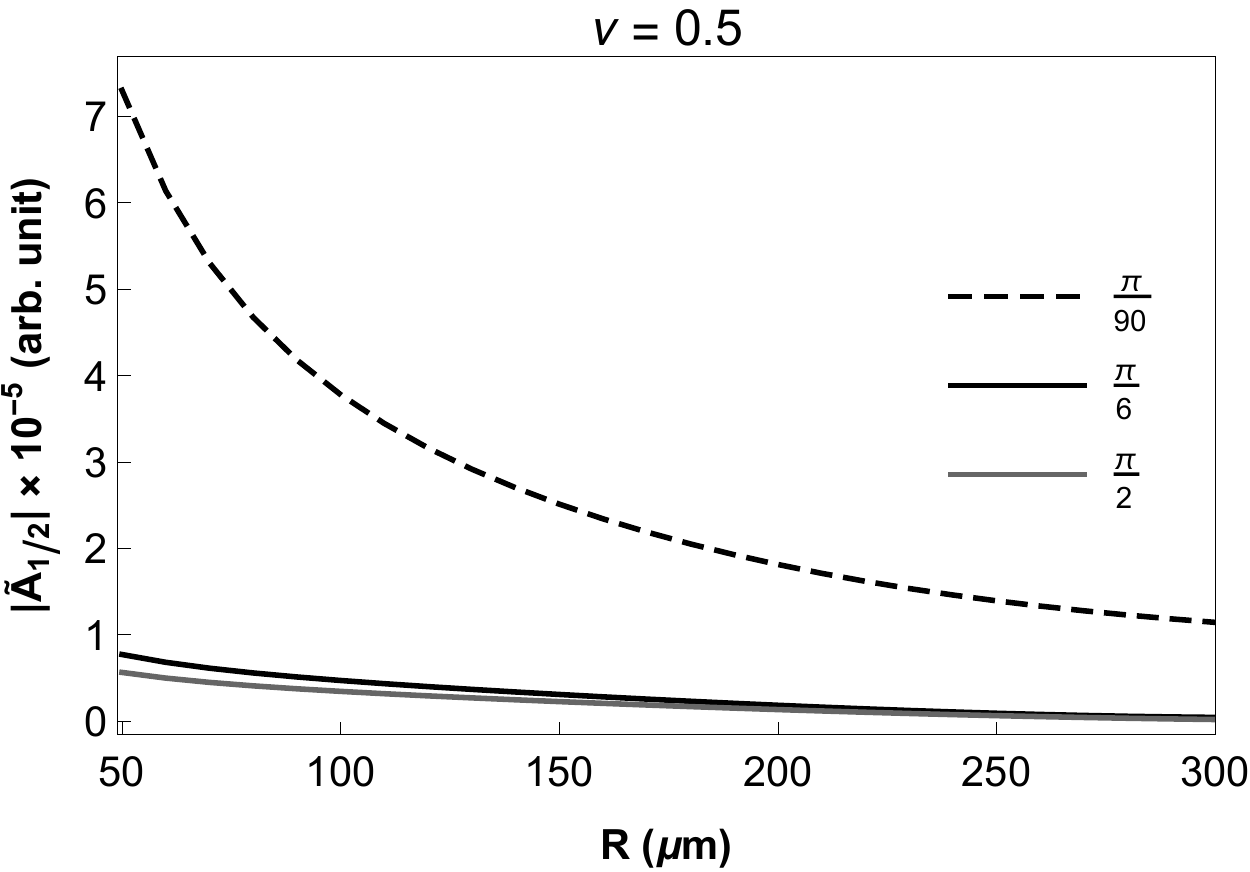}
\caption{R-dependence of radiation potential given by Eq.(\ref{anis}) for different angles. Here we take the observation point on the cutting boundary $R\equiv R_1$.}
\label{fig.3}
\end{center}
\end{figure}
  As it follows from Fig.3 at large distances the potential behaves as $\sim 1/R$. However, the amplitude is different for different observation angles in contrary to the case in the previous paragraph.

  The radiation potential in the lossless medium is highly anisotropic. It follows from Fig.3 that at large observation angles the potential is significantly smaller. Maximum potential is reached at small angles from the particle trajectory. Modifying the imaginary part of $n_0$, one can observe a transition from highly directed to isotropic radiation potential.  As it is shown in the previous paragraph, isotropic radiation potential leads to dipole like radiation intensity. However, in the lossless medium the radiation potential is highly directed.
\begin{figure}
\begin{center}
\includegraphics[width=8.0cm]{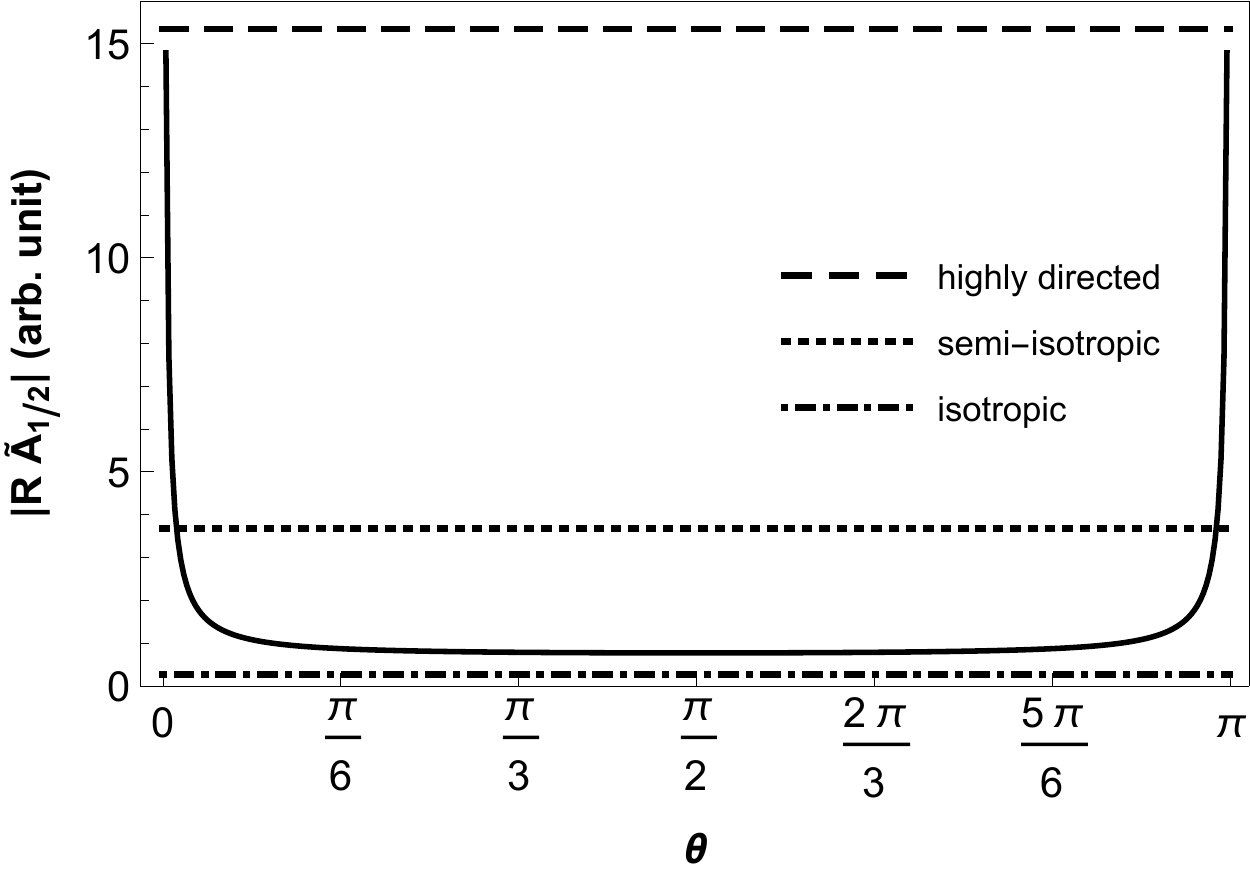}
\caption{Angular distribution of radiation vector potential.  Straight lines are the potentials for the media with losses for different values of refraction index imaginary part }
\label{fig.4}
\end{center}
\end{figure}
Note that the $R$-dependence for all angles is $\sim 1/R$, see Fig.3 . Straight lines in Fig.3 represent the radiation potentials  for different values of $Im[n_0]$ calculated via Eq.(\ref{vecpotspec}). The $U$-shape curve is obtained for the lossless medium using generalized Green's function Eq.(\ref{grfuncother},\ref{anis}). The actual radiation potential is determined by the curve below the straight lines.  Therefore one can distinguish three different regimes of radiation depending on the losses of the medium ($Im[n_0]$), see Fig.4.

Physically, the above-mentioned transition from one angular distribution to another can be understood in the following way. Vector potential of a moving charge in the vacuum behaves like $A_0(R)\sim K_0(\sqrt{k_0^2-k^2}R|\sin\theta|)$. This means that it is concentrated around the direction of the particle's velocity because for large arguments $K_0$ falls exponentially. This alone is not a photonic field yet. When one adds a nonhomogeneous medium these pseudophotons scatter becoming real photons. In the Maxwell fish eye medium without losses, photons are totally reflected from layers with decreasing refraction index. Because of the total internal reflection, those remain around the charge. When the losses are taken into account photons evanescently penetrate through the layers and eventually result in the isotropic angular distribution for large enough losses as seen in Fig.3. This transition is similar to the phenomenon of attenuated total reflection (ATR), see, for example, \cite{atr}.
 \section{Conclusion and discussion} We have considered the radiation from a charged particle moving through a medium with Maxwell fish eye refraction index profile. The spectrum of radiation has discrete character. The main emitted wavelength is proportional to the radius of refraction profile $\lambda=4\pi\sqrt{\varepsilon/3}\rho$. In the regular medium (with losses) the radiation has dipole character, whereas in the lossless medium it is highly directed. In the intermediate regime with moderate losses radiation will be non-isotropic.
 %Cherenkov condition is stronger than corresponding one for homogeneous medium with refraction index as in origin and is weaker compared to the %condition for medium with refraction as at radius.

  So far we assumed that particle trajectory along $z$ passes through the origin. If the trajectory is at some distance $d$ from the origin then the corresponding current density is determined as
\begin{equation}
{\bf j}({\bf r},\omega)=e\frac{{\bf v}}{v}\delta(x-d_x)\delta(y-d_y)e^{ik_0z}
\label{impact}
\end{equation}
Similar calculations show that all expressions keep their form except $\rho$ in the argument of the Bessel function Eq.(\ref{inten}). It should be substituted by $\sqrt{\rho^2+d^2}$ where $d^2=d_x^2+d_y^2$. So for the main emitted wavelength $\lambda=4\pi\sqrt{\varepsilon/3}\rho$, one has
\begin{equation}
I_{1/2}^d(\theta)=\frac{3e^2}{\pi c}\frac{K_0^2(\frac{\sqrt{\frac{3}{4\varepsilon}+k^2d^2}}{\beta})}{\sinh^2(\frac{3\pi Im\varepsilon}{8\varepsilon})}sin^2\theta
\label{intendis}
\end{equation}
Correspondingly, Cherenkov condition acquires the form
\begin{equation}
\beta>\frac{\sqrt{3}}{2\sqrt{\varepsilon}}\sqrt{1+\frac{d^2}{\rho^2}}
\label{cherend}
\end{equation}
This condition means that the impact distance should be smaller than the emitted wavelength $d<\lambda/2\pi$.

 In the anisotropic case radiation potential depends not only the ratio $d/\lambda$ as in isotropic case but also $d/\rho$. Therefore here restriction on the impact parameter is stronger $d/\rho\ll 1$.

Note that Maxwell fisheye millimeter scale systems already exist in 2D \cite{2D} as well as in  3D \cite{3D}. Therefore they can be used for the generation of radiation in microwave and terahertz regions. As it seen from Eq.(19), radiation intensity does not depend on the cutting parameter $R_1$ (except the lossless medium case). This means that real systems with Maxwell fish eye profile can have a size of order radius $\rho$ \cite{2D,3D}.
  %This equation could has  either one or three real solutions. Respectively, depending on the value of $\kappa, n_0, s$  we could get one or three %profiles which admit perfect imaging
  %in the presence of polarized light. We are planning to  discuss these issues  elsewhere.

%Then we extend our consideration to the case of spinning light, i.e. includes %light polarization.

\acknowledgements
Authors are  grateful to Armen Allahverdyan,  and Arsen Hakhoumian  for useful discussions and comments. This work was performed within partial financial support from Armenian Committee of Science Grant  18T-1C082 (Zh.G.).


\begin{thebibliography}{99}
%\bibitem{Pendry06} J.Pendry, D. Schurig, and D.Smith, %\emph{Controlling electromagnetic fields}, Science {\bf 312}, 1780,
% 1780-1782, (2006).
\bibitem{bowolf} M.Born and E.Wolf,\emph{Principles of Optics},Fourth Edition,Pergamon Press,(1970).
\bibitem{Leonhardt06} U.Leonhardt, Science {\bf 312}, 1777, %1777 - 1780,
 (2006);  New J. Physics,{\bf 8}, 118, (2006).
%\bibitem{lai09}Y.Lai, H.Chen, Zh.-Q.Zhang, and C.T.Chan,  Phys.Rev.Lett. {\bf 102}, 093901 (2009).
%\bibitem{gbur13}G.Gbur,Progress in Optics,{\bf 58}, 65, (2013).
%\bibitem{zhou05} J.Zhou, T.Koschny, M.Kafesaki, E.N.Economou, J.B.Pendry, C.M.Soukolis, Phys.Rev.Lett., {\bf 95}, 223902 (2005).
%\bibitem{Sun08}J.Sun, Ji Zhou  and L.Kang, Optics Express, {\bf 16}, 17768, (2008).
%\bibitem{chen11} X.Chen, Yu Luo, J.Zhang, K.Jiang, J.B.Pendry and S.Zhang,
%Nature Communications, DoI: 10.1038/ncomms1176.
%\bibitem{chen13}H.Chen, B.Zheng, L.Shen, H.Wang, X.Zhang, N.I.Zheludev and B.Zhang, Nature Communications, DOI: 10.1038/ncomms3652
%(2013).
%\bibitem{choi14} J.S.Choi and J.C.Howell,  Optics Express, {\bf 22 (24)}, 29465-29478, (2014).
%\bibitem{Maxwell}J.C.Maxwell,  Camb. Dublin Math. J. {\bf 8}, 188, (1854).

 \bibitem{Leonhardt09} U.Leonhardt,  New J. Phys. {\bf 11}, 093040-093051 (2009).
  \bibitem{Philbin10} U.Leonhardt and T.G.Philbin,Phys. Rev. A {\bf 81}, 011804 (2010).
  \bibitem{Blaikie10} R.J. Blaikie,  New Journal of Physics ,{\bf 12}, 058001 (2010).
  \bibitem{pazynin15} Leonid A. Pazynin, Vadim L. Pazynin, and Hanna O. Sliusarenko,IEEE TRANSACTIONS ON ANTENNAS AND PROPAGATION, {\bf 63},4393,(2015), doi:10.1109/TAP.2015.2465832.
  \bibitem{perczel2018}J. Perczel, P. K$\acute{o}$m$\acute{a}$r, and M. D. Lukin, Phys.Rev.A,{\bf 98}, 033803 (2018).
\bibitem{turks14} K.Dadashi,H.Kurt,K.Ustun and R.Esen,  J.Opt.Soc.Am.B, {\bf 31}, 2239-2245, (2014).
\bibitem{gevdav20}Zh.Gevorkian, M.Davtyan and A.Nerssesian, Phys.Rev.A, {\bf 101},023840, (2020).
\bibitem{singap}Yangjie Liu and L. K. Ang, Scientific Reports,{\bf 3} : 3065, (2013). | DOI: 10.1038/srep03065.
\bibitem{dyadic}C.-T. Tai, Differential operators in vector analysis and the Laplacian of a vector in
the curvilinear orthogonal system. Tech. Rep., Radiation Laboratory, University
of Michigan, Ann Arbor, MI (1990);C.-T Tai,  Dyadic Green Functions in Electromagnetic Theory. The IEEE PRESS
Series on Electromagnetic Waves (IEEE PRESS, 1993), 2nd edn.

\bibitem{demkov71}Yu. N. Demkov and V. N. Ostrovskii,SOVIET PHYSICS JETP, {\bf 13},1083,(1971).
\bibitem{poland}Radosław Szmytkowski, J. Phys. A: Math. Theor. {\bf44},065203 (2011).
\bibitem{pazynin}L.A. Pazynin and G.O. Kryvchikova,Progress In Electromagnetics Research,  {\bf 131}, 425-440, (2012).
\bibitem{bound}W. C. Chew,arXiv:1406.4780v1,(2014).
\bibitem{ryzhik} I.S. Gradshteyn and I.M. Ryzhik, \emph {Table of integrals, series, and products}, Academic Press,(2007).

\bibitem{mik72} M.L.Ter-Mikaelian, High-Energy Electromagnetic Processes in Condensed Media,
N.Y., John Wiley and Sons Inc.,(1972).
\bibitem{atr} P.Larkin, Infrared and Raman Spectroscopy: Principles and Spectral Interpretation, Elsevier (2011); https://doi.org/10.1016/C2010-0-68479-3.
\bibitem{2D}Daniel Headland,Masayuki Fujita and Tadao Magatsuma,Optics Express, {\bf 28}, 2366    (2020).
\bibitem{3D}Hui Feng Ma, Ben Geng Cai, Teng Xiang Zhang, Yan Yang, Wei Xiang Jiang, and
Tie Jun Cui,IEEE Transaction on Antennas and Propagation,  {\bf 61},2561,(2013).

\end{thebibliography}
\end{document}